\documentclass[12pt]{article}
\usepackage{amsmath,epsfig,graphicx,amssymb}
\usepackage[margin=0.7in]{geometry}
\newcommand{\beq}{\begin{equation}}
\newcommand{\eeq}{\end{equation}}
\newcommand{\ben}{\begin{eqnarray}}
\newcommand{\een}{\end{eqnarray}}
\newcommand{\bes}{\begin{subequations}}
\newcommand{\ees}{\end{subequations}}
\newcommand{\bFig}{\begin{figure}}
\newcommand{\eFig}{\end{figure}}
\date{}
\begin{document}
\title{The Unfinished Search for Wave-Particle and Classical-Quantum Harmony}
\author{Partha Ghose\footnote{partha.ghose@gmail.com} \\
The National Academy of Sciences, India,\\ 5 Lajpatrai Road, Allahabad 211002, India.}
\maketitle

\begin{abstract}
The main purpose of this paper is to review the progress that has taken place so far in the search for a single unifying principle that harmonizes (i) the wave and particle natures of matter and radiation, both at the quantum and the classical levels, on the one hand and (ii) the classical and quantum theories of matter and radiation on the other hand. In the author's opinion, the Koopman-von Neumann-Sudarshan (KvNS) Hilbert space theory based on complex wave functions underlying particle trajectories in classical phase space, is an important step forward in that direction. To appreciate the similarities and differences between classical and quantum wave functions, it is important first to review the famous paradoxes of quantum theory arising mainly from the dual character of matter and radiation and the mysterious nature of measurements in quantum theory. They have given rise to the suspicion that quantum mechanics is an incomplete theory and to multiple interpretations of quantum mechanics as well as no-go theorems ruling out certain types of hidden variable theories introduced to `complete' quantum mechanics in some way. It is also important to point out that experimental verifications of the predictions of the double-prism experiment and the observation of average single photon trajectories in weak measurements appear to favour the spacetime picture of particles favoured by Einstein, de Broglie and Bohm over the complementarity approach favoured by Bohr. The KvN theory of classical mechanics provides a clear and beautiful harmony of classical waves and particles. Sudarshan has given an alternative but equivalent formulation that shows that classical mechanics can be regarded as a quantum theory with essentially hidden noncommuting variables. An extension of KvNS theory to classical electrodynamics provides a sound Hilbert space foundation to it and satisfactorily accounts for entanglement and Bell-CHSH-like violations already observed in classical polarization optics. An important new insight that has been obtained through these developments is that entanglement and Bell-like inequality violations are neither unique signatures of quantumness nor of non-locality---they are rather signatures of non-separability. Another new insight is the interpretation of the Wigner function as a KvNS wave function, i.e. a probability amplitude which need not be positive everywhere. This has important implications for simulating certain types of quantum information processes using classical polarization optics. Finally, Sudarshan's proposed solution to the measurement problem using KvNS theory for the measuring apparatus is sketched to show to what extent wave and particles can be harmonized in quantum theory. 

\end{abstract}
\tableofcontents
\section{Introduction}
The nature of the wave function in quantum mechanics has been debated from the the very beginning. Paradoxes like wave-particle duality, the mystery of quantum measurements and conundrums such as non-locality, have spawned a plethora of interpretations of quantum mechanics \cite{gen} and led to the belief that wave functions are special to quantum mechanics. Wave functions span a Hilbert space, and generations of text books have given rise to the belief that wave functions and Hilbert spaces occur only in quantum physics. One of the main purposes of this paper is to show that they can be introduced also in classical physics. In fact, this was done way back in the early 1930s by Koopman \cite{K} and von Neumann \cite{vN}, and later by Sudarshan \cite{sud} who had a different but equivalent approach. Koopman showed that the phase space of a classical mechanical system can be converted into a Hilbert space by postulating a scalar product which is an integration rule over the points of the phase space. This inspired von Neumann to apply the formalism to the ergodic problem. The dynamics in phase space is described by a classical probability density constructed from an underlying wave function--the Koopman-von Neumann (KvN) wave function. This is analogous to the Born rule in quantum mechanics. In the KvN framework, observables are represented by commuting self-adjoint operators acting on the Hilbert space of KvN wave functions, ensuring that all observables are simultaneously measurable. This is in contrast to quantum mechanics in which the hermitian operators representing observables need not all commute and cannot all be measured simultaneously with precision. A significant new dimension was added to the KvN theory by Sudarshan \cite{sud} who showed how classical mechanics can be embedded in quantum mechanics with essentially {\em hidden variables}, and how a {\em superselection rule} operates to rule out transitions between classical states with different phases, thus causing decoherence. He further showed how KvNS theory provides the appropriate theoretical framework to solve the measurement problem in quantum theory. All this has led, in the author's view, to significant progress {\em towards} the search for a unifying principle that harmonizes waves and particles as well as classical and quantum theories.

An unexpected recent discovery is that classical polarization optics displays quantum-like features such as entanglement, as originally predicted by Spreeuw \cite{sp} and independently by Ghose and Samal \cite{g}. This is now an emerging field, and recent developments \cite{radial, oron, kozawa, souza, holleczek, borges, simon, gabriel, eberly,  agarwal, kagalwala, ghose2, pereira} have been reviewed by Ghose and Mukherjee \cite{ghose}. Aiello and colleagues have developed a unified theory for different kinds of light beams exhibiting classical entanglement and have indicated several possible extensions of the concept \cite{aiello}. It has also been recently shown that the origin of such apparent quantumness in patently classical phenomena indeed lies in the KvNS Hilbert space structure of classical electrodynamics \cite{raja}.

Another very significant result is the demonstration by Bondar and colleagues \cite{bondar, bond2} that the Wigner function is proportional to the KvNS wave function which is a probability amplitude and need not be positive everywhere, thus solving the problem of its mysterious negative values in spite of being a probanility density. This makes it possible to use classical polarization optics to simulate certain types of quantum information processes.

I will first briefly sketch the principal quantum paradoxes, interpretations of quantum mechanics and the no-go theorems, and then touch upon the epistemic versus ontic character of the quantum wave function before introducing wave functions in classical physics. This will enable the reader to see more clearly unexpected similarities between classical and quantum theories and help in identifying the real differences between them. In particular, it will become evident that entanglement and consequent violations of Bell-CHSH type inequalities are common features of classical optics and quantum mechanics, and that realism and separability are sufficient for the derivation of Bell's theorem, not locality. I will then briefly refer to the violation of non-contextuality in classical polarization optics and its implications for classical realism. I will end with some remarks on the unity of waves and particles in the KvNS theory of classical mechanics, and sketch Sudarshan's demonstration that the quantum measurement problem is solved by using a certain representation of KvNS theory for the classical measuring apparatus. 

It is assumed that the reader has a sound knowledge of the basics of classical and quantum mechanics and classical electrodynamics.

\section{Principal Quantum Paradoxes}
{\flushleft{\em Wave-Particle Duality}}
\vskip0.1in
When a single particle state is incident on a double-slit, it produces an interference pattern on a distant plate when both slits are open, and a single-slit diffraction pattern when one of the slits is closed. There are two puzzles here: the first one is that an interference pattern is seen at all with `particles', and the second one is that this pattern disappears when one tries to figure out which slit the particle actually went through. Heisenberg's uncertainty principle $\Delta x \Delta p \geq \hbar/2$ plays a key role here by ruling out precise particle trajectories between the source and the detector and also introducing untrollable disturbances in measurements that wash out the relative phases responsible for the interference pattern \cite{feynman}. This is the essence of the paradox of wave-particle duality. This paradox has led to a number of interpretations ranging from particles {\em or} waves as expressed by Bohr in his {\em complementarity principle} \cite{bohr} through {\em which way} experiments \cite{rauch1, rauch2, rauch3} and their interpretations in terms of partially particle-like and partially wave-like `information' \cite{ wooters, greenberger, englert, ghose b} to fully particle {\em and} wave interpretations \cite{dbb, gha, gh} and their tests \cite{mo, brida}.

The double-prism experiment \cite{gha} to probe the nature of wave-particle duality differs from the others in that, instead of using interference as a signature of a wave-like property, it uses {\em tunneling} as the signature of a wave-like property on the one hand and anti-coincidence on a beam splitter of {\em indivisible} quanta as a signature of a particle-like property on the other hand. Quantum mechanics predicts both tunneling and anti-coincidence of single photons (Fig. 1). Since every single photon is either reflected or transmitted, the {\em same} experimental set up exhibits {\em both} fully particle-like and fully wave-like characteristics, thus favouring the wave {\em and} particle interpretation over the wave {\em or} particle interpretation.
\begin{figure}
{\includegraphics[scale=0.5]{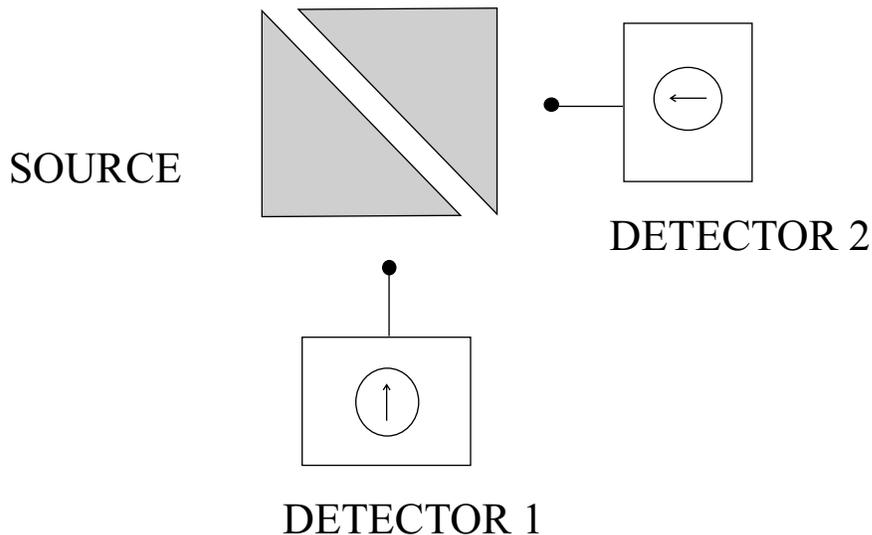}}
\caption{\label{Figure 1}{\footnotesize The principle of the double-prism experiment proposed by Ghose, Home and Agarwal (Ref. \cite{gha}). Single photons are incident from the left, and are either totally internally reflected to detector 1 or tunnels through to detector 2. The two detectors are predicted to click in anticoincidence.}}
\end{figure} 

In 2001 Cufaro-Petroni and Vigier \cite{vigier} analyzed various single-neutron and single-photon experiments as well as their theoretical interpretations in terms of the de Broglie-Bohm theory and Nelson's stochastic mechanics \cite{nelson}. Vigier emphasized that though no conclusive argument could be reached, they seemed to favour Einstein's and de Broglie's realistic spacetime description of light in terms of particle trajectories. 

\vskip 0.1in
{\flushleft{\em The Schr\"{o}dinger Cat}}
\vskip 0.1in
The cat paradox was introduced by Schr\"{o}dinger in his 1935 paper \cite{schr} on the status of the quantum theory. He showed that the uncertainties of the microscopic world (the uncertainty of the time of decay of a radioactive substance) can get transferred to a macroscopic object like a cat, resulting in a superposition of incompatible macroscopic states like `dead' and `alive' which are never seen. Interestingly, following the work of Leggett and his co-workers \cite{leg1, leg2, leg3}, recent experiments have successfully created mesoscopic cat states using rf-SQUIDs \cite{squid}.

\vskip 0.1in
{\flushleft{\em The EPR Paradox}}
\vskip 0.1in
In the same paper Schr\"{o}dinger dwelt extensively with the idea of the non-separability of two quantum systems that are spatially separated and non-interacting, introduced by Einstein, Podolski and Rosen \cite{epr}. He coined the term `entanglement' and argued that it is {\em the} characteristic feature of quantum mechanics that distinguishes it from classical physics. Entanglement is the loss of identity of individuals in a holistic entity, the entangled state. When a measurement is made on this holistic entity in a particular basis, it results in a `conditional disjunction' of this entity by an obscure process outside the ambit of quantum theory. This leads to the EPR paradox and the incompleteness argument. The famous EPR paper was written by Podolski and, according to Einstein, the main point of the argument got obscured by pedantry. Einstein's argument is very simple \cite{ein}. Let $S_1$ and $S_2$ be two partial systems, and let 
\ben
\Psi(S_1 S_2) &=& c_1 \psi_1(S_1) \psi_2(S_2) + c_2 \psi_2(S_1) \psi_1(S_2)\nonumber\\
&=& c^\prime_1 \phi_1(S_1) \phi_2(S_2) + c^\prime_2 \phi_2(S_1) \phi_1(S_2) \label{EPR}
\een
with $|c_1|^2 + |c_2|^2 = |c^\prime_1|^2 + |c^\prime_2|^2 =1$ be their wave function expressed in two incompatible bases. Such a wave function is non-separable in the sense that it is not expressible as the product of the individual wave functions of $S_1$ and $S_2$. This results in $S_1$ and $S_2$ losing their individualities in the state $\Psi(S_1 S_2)$. It is in this sense that they become `non-separable' or `entangled'. This is a straightforward consequence of the fact that quantum mechanical wave functions span Hilbert spaces. 

Einstein assumes that an experimenter is free to choose the basis in which a measurement is to be made on $S_1$. Then it is clear that she would obtain different wave functions $\psi(S_2)$ or $\phi(S_2)$ for $S_2$ depending on her choice. Now, if one further assumes that (i) the ontic state of $S_2$ is not in any way influenced by what measurement is made on $S_1$ (locality/separability assumption), and that (ii) every ontic state of a system is associated with a single wave function (a requirement of standard quantum theory), then there is a problem. One way to avoid this problem is to allow more than one wave function to be associated with a given ontic state of a system, i.e. to give up condition (ii). In this case the wave function may be called `epistemic' (i.e. more than one wave function correspond to a given ontic state), and quantum mechanics is clearly an `incomplete' description of nature. Einstein favoured this interpretation. The alternative is to avoid assumption (i), i.e. the assumption of locality/separability. The notions of locality and separability are, however, not identical--it is possible to have non-separability without non-locality, as we aill see later when discussing classical polarization optics. Einstein found both choices unacceptable.

In the 1927 Solvay Conference Einstein had already sketched a simple argument to show that even in the case of a single particle, quantum mechanics was incompatible with locality and realism \cite{bac}. This was a precursor to Bell's theorem \cite{bell1}. Consider a single particle represented by a plane wave function incident on a screen with a small hole. On the other side of the hole, the wave function spreads out in the form of a spherical wave, and is finally detected by a large hemispherical detector. This spherical wave obviously shows no preferred direction. Einstein observed:
\begin{quote}
If $|\psi|^2$ were simply regarded as the probability that at a certain point a given particle is found at a given time, it could happen that {\em the same} elementary
 process produces an action in {\em two or several} places on the screen. But the interpretation, according to which the $|\psi|^2$ expresses the probability that
 {\em this} particle is found at a given point, assumes an entirely peculiar mechanism of action at a distance, which prevents the wave continuously distributed in 
space from producing an action in {\em two} places on the screen.
\end{quote}
Einstein noted that this `entirely peculiar mechanism of action at a distance' was in contradiction with the postulate of relativity
\section{Interpretations of Quantum Mechanics}
\vskip 0.1in
{\flushleft{\em Bohr}}
\vskip 0.1in
As a result of these paradoxes, quantum mechanics has been riddled with interpretations of various kinds over the years. The first interpretation given by Niels Bohr was an epistemological one \cite{bohr}. Bohr emphasized that since all measurements must result in unambiguous results, one must necessarily use a {\em classical} measuring apparatus that is free from quantum mechanical uncertainties. The system and the apparatus enter into an `unanalyzable whole' from which only statistically valid results can be inferred by a process that is left unspecified. The questions that naturally arise are: (i) Where is the cut between the classical and the quantum domains? (ii) Can one be more specific about the `unanalyzable whole'? These questions have engaged physicists and philosophers ever since in what appears to be an interminable debate. 
\vskip 0.1in
{\flushleft{\em von Neumann}}
\vskip 0.1in
If the quantum world is taken to underlie the classical world (all apparatuses, after all, are made of atoms and molecules which are known to behave quantum mechanically) and if quantum mechanics (as a physical theory) is to have universal validity, Bohr's approach appears unsatisfactory. Furthermore, the dynamical evolutions of quantum and classical systems follow different laws, and hence there is no consistent way of coupling them, and therefore of using classical systems to measure quantum systems. These factors were taken into account by von Neumann \cite{vN2} by treating the measuring apparatus quantum mechanically. This, however, resulted in an entangled or non-separable state $\rho(S,A)$ of the system S and the apparatus A, and he had to introduce projection operators $\Pi_i=|A\rangle_i\langle A|_i$ (where $|A\rangle_i$ are the complete set of final {\em classical} states of the apparatus) to reduce this pure state to a mixed state $\hat{\rho}(S,A) = \sum_i \Pi_i \rho(S,A)\Pi_i$ with $Tr \hat{\rho}(SA) = Tr \rho(SA) = 1$ but $\hat{\rho}(SA)^2 \neq \rho(SA)$. The elements of this diagonal matrix represent the probabilities $p_i =|c_i|^2$ of the various measurement outcomes. A particular final state $|A\rangle_i$ of the apparatus corresponds to a particular state $|S\rangle_i$ of the system. Thus, `if this, then that' is what can be said, and the final state is a product state $|S\rangle_i|A\rangle_i$ with probability $p_i$. This `conditional disjunction' of an holistic state having coherence into distinguishable individuals lacking mutual coherence is non-unitary, and hence it must lie outside the domain of quantum mechanics, and consequently outside the domain of any physical theory if quantum mechanics is to have universal validity in the physical world. This process of measurement was dubbed `process 1' by von Neumann. It is as mysterious as the Bohr process. He termed the unitary Schr\"{o}dinger evolution of quantum systems `process 2'. It is `process 1' that most interpretations have tried to avoid or eliminate.
{\flushleft{\em The Many-Worlds Interpretation}}
\vskip 0.1in
 In order to avoid `process 1' Hugh Everett proposed the `relative state' interpretation \cite{ev} which later on came to be transformed into the `many-worlds' interpretation \cite{dewitt}. He achieved this by introducing a process of `splitting' of the wave function into branches all of which exist simultaneously. This implies that all possible outcomes of a measurement actually exist {\em simultaneously}, and no collapse really takes place. This means that a cat is both alive and dead, but these two states are in different branches of the wave function which do not interact with each other. The main problem with this interpretation is that it is difficult to define probabilities in it \cite{hargreaves}. Since all `possible' outcomes actually occur, it appears meaningless to talk of probabilities for branches (other than 0 and 1). Even if it does make sense to talk of nontrivial probabilities for branches, it is not clear what ensures that the probabilities in a many-worlds interpretation agree with those of standard quantum mechanics \cite{adlam, dawid}. Nevertheless, Aguirre and Tegmark \cite{teg} have proposed a cosmological version of the many-worlds interpretation. 

\vskip 0.1in
{\flushleft{\em Hidden Variable Theories}}
\vskip 0.1in
The EPR argument that quantum mechanics is incomplete spawned a variety of hidden variable theories whose aim was to complete quantum mechanics \cite{gen2}. The basic idea was to introduce `hidden variables' underlying the observables in such a way as to restore realism, and often also determinism, at the fundamental level, and yet recover the quantum mechanical predictions as averages over these hidden variables. However, most of these theories were subsequently shown to be incompatible with quantum mechanics (and hence really different theories) by certain no-go theorems, as we will shortly see, except the one proposed by Bohm (and earlier by de Broglie) \cite{dbb}. In the de Broglie-Bohm theory, the position of a particle is introduced as an ontic hidden variable, and it is piloted by the wave function which is a solution of the Schr\"{o}dinger equation. Once the distribution $P(x,t_0)$ of particles is matched with the quantum mechanical distribution $|\psi(x,t_0)|^2$ at some intial time $t_0$, a continuity equation guarantees that it agrees with $|\psi(x,t)|^2$ at all future times $t$, ensuring that the ensemble averages of observables calculated from the distibution agree with the corresponding expectation values in quantum mechanics at all times. The theory is therefore completely equivalent to quantum mechanics in its observable predictions, but it restores realism (and also determinism) at the fundamental ontic level at the expense of locality. 
The theory is explicitly non-local because the position and motion of each particle generally depends on the coordinates of the whole system. In fact, Bohm often emphasized the non-locality or wholeness inherent in this interpretation as the only real significance of quantum physics vis-a-vis classical physics \cite{hiley}. One would have expected Einstein to reject this interpretation on that ground alone, but strangely, he objected to it for its determinism. In a letter to Born dated 12 May, 1952 \cite{eins2}, he wrote:
\begin{quote}
Have you noticed that Bohm believes (as de Broglie did, by the way, 25 tears ago) that he is able to interpret the quantum theory in deterministic terms? That way seems too cheap to me. But you, of course, can judge this better than I.
\end{quote}
A comprehensive account of the causal theory of Bohm will be found in the book by P. R. Holland \cite{holland}.

The matching of the probability distributions $P(x,t) = |\psi(x,t)|^2$, known as the {\em quantum equilibrium hypothesis}, is a matter of some concern which has been addressed by D\"{u}rr and his colleagues \cite{durr} and Valentini who has shown that superluminal signalling would be possible without this hypothesis \cite{valentini}. 
\begin{figure} [h]
{\includegraphics[scale=1.0]{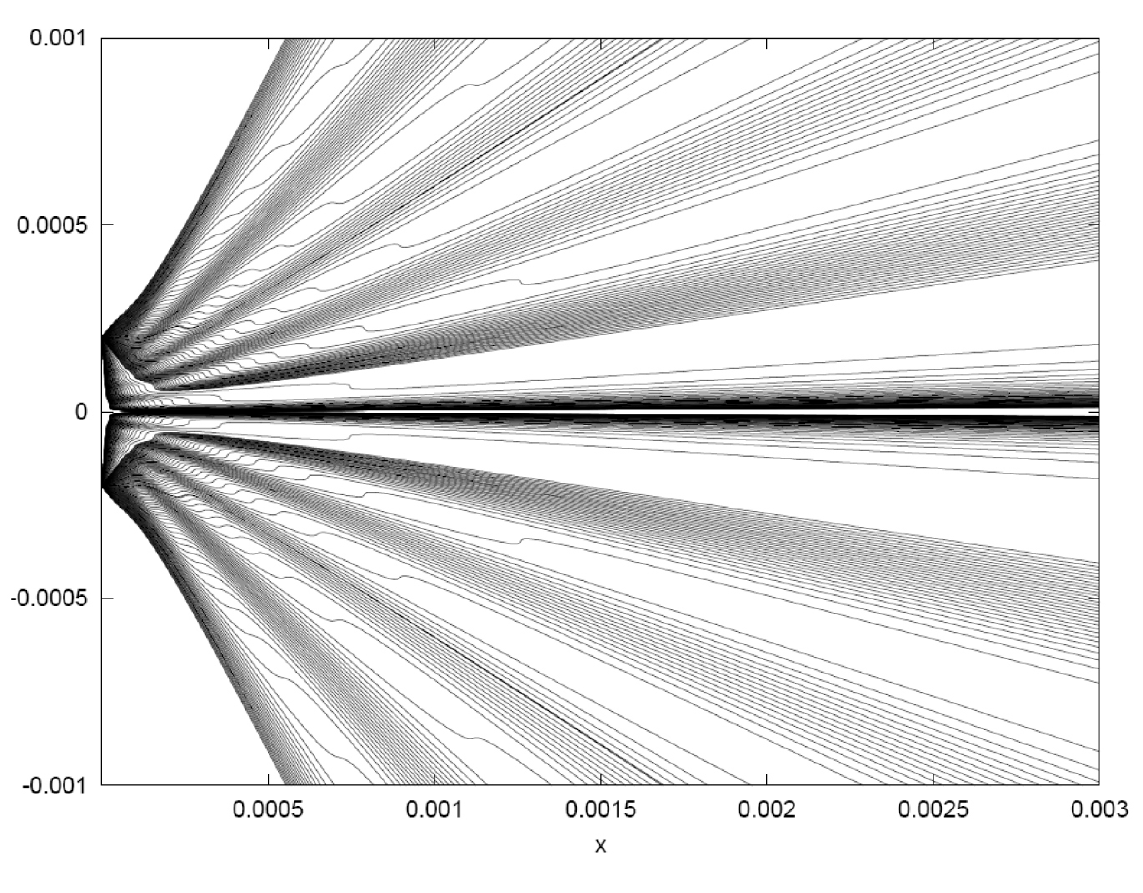}}
\caption{\label{Figure 2}{\footnotesize Bohmian trajectories of photons calculated by Ghose, Majumdar, Guha and Sau (Ref. \cite{gmg}).}}
\end{figure}
\begin{figure} [h]
{\includegraphics[scale=1.0]{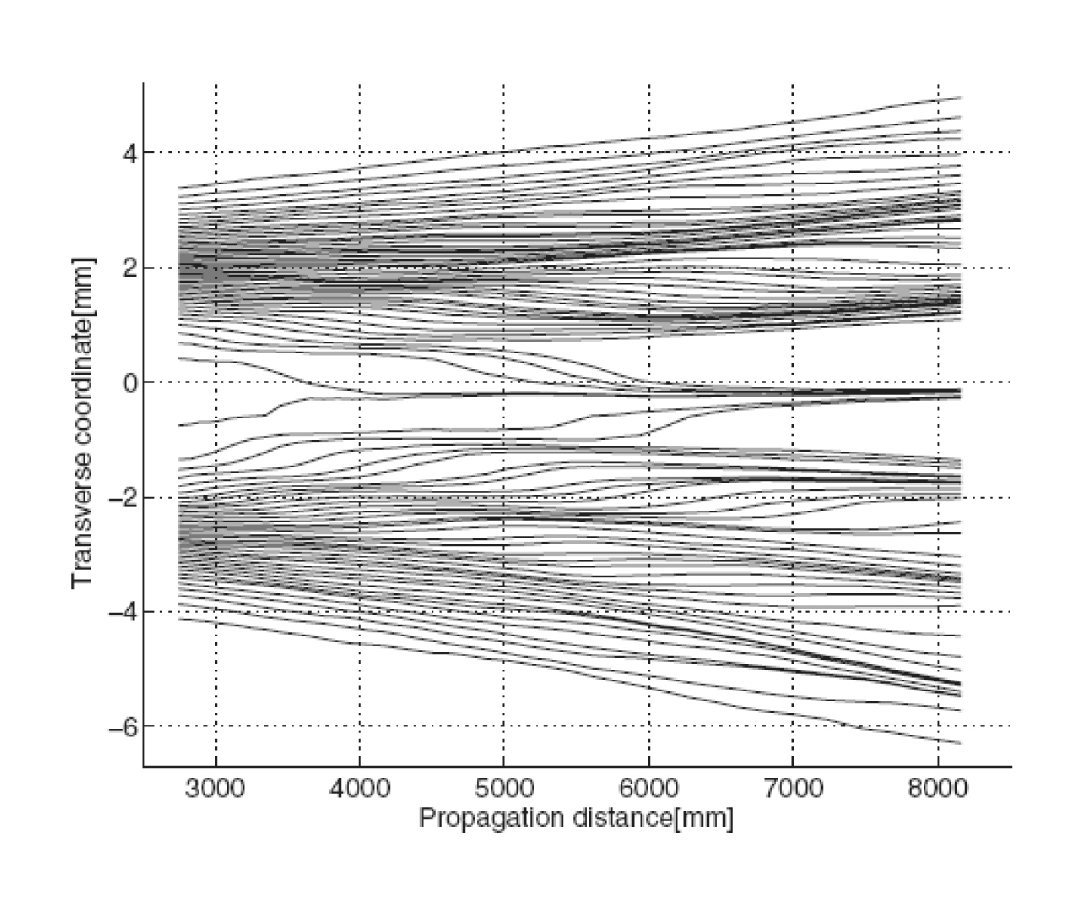}}
\caption{\label{Figure 3}{\footnotesize Average trajectories of single photons observed in weak measurements by Steinberg and colleagues (Ref. \cite{steinberg}).}}
\end{figure}
The matching at all times for entangled or non-separable particles turns out to be problematic \cite{ghose3, marco2}, requiring a specific choice of the relative orientation of the configuration spaces of the particles, which is not dictated by the theory itself. 

Initially, it was considered impossible to define boson trajectories in the de Broglie-Bohm theory in view of the difficulties of describing bosons relativistically \cite{hiley}. During 1993-1996, Ghose and his colleagues developed a relativistic quantum mechanical description of spin-0 and spin-1 bosons starting from the Kemmer-Duffin-Petiau (KDP) equation \cite{kemmer} and the Harish-Chandra equation for massless bosons \cite{hc}, laying the foundations for a de Broglie-Bohm theory of such bosons \cite{ghosehome}. They computed Bohmian trajectories of photons for specific cases in 2001, using the Harish-Chandra equation \cite{gmg} (Fig. 2). Subsequently in 2009, experiments by Steinberg and his colleagues \cite{steinberg} using techniques of weak measurements \cite{weak} showed trajectories (Fig. 3) qualitatively similar to the predicted ones. However, weak measurements have been the subject of some controversy \cite{cont}.

{\flushleft{\em Other Interpretations}}
\vskip 0.1in

There are some other interpretations of quantum mechanics such as the consistent histories interpretation \cite{griffiths}, the transactional interpretation \cite{cramer}, the modal interpretation \cite{modal} and relational quantum mechanics \cite{rov}, details of which will be found in \cite{gen}. The most recent interpretation is Quantum Bayesianism \cite{fuchs} which uses an informational or epistemic interpretation of the wave function and views collapse as updating of information on obtaining new information, and hence not disturbing the ontic state in any manner. It is free of most of the paradoxes and conundrums of quantum theory.

\section{No-Go Theorems for Hidden Variables}

Some no-go theorems were proven later to show that most hidden variable theories were inconsistent with quantum mechanics, and hence presumably empirically false too. The first one of these was the von Neumann theorem. 
\vskip 0.1in
{\flushleft{\em The von Neumann Theorem}}
\vskip 0.1in
What von Neumann attempted was to prove that dispersion free states, i.e. hidden variables, were impossible. He first noted that any real linear combination of any two Hermitian operators represents an observable, and then made use of the further assumption that the same linear combination of expectation values was the expectation value of the combination. This assumption is certainly true of quantum mechanical states, but not of all dispersion free states, as was first shown by Bohm and later by Bell \cite{bell2}. Alhough von Neumann's assumption was applicable to a class of hidden variables, it was not applicable to hidden variables of the Bohm type which were `contextual' in the sense that the observables in the theory were not properties of the observed system alone but also of the measuring apparatus. 

Other associated theorems like that of Jauch and Piron and of Gleason, and their limitations, will be found in Bell's paper \cite{bell2}. 
{\flushleft{\em Bell's Theorem}}
\vskip 0.1in
Bell proved his famous theorem \cite{bell1} in 1964, showing that quantum mechanics was incompatible with local realism. The readers will find an engaging review of Bell's philosophy and work, particularly Bell's theorem and its experimental verification, in Bertlmann's recent paper \cite{bert}. It is worth noting that like Einstein, Bell too was disturbed by the non-locality implied by his theorem and its experimental confirmation. As reported by Bertlmann, 

\begin{quote}
The nonlocality feature disturbed John deeply since for him it was equivalent to a {\em `breaking of Lorentz invariance'} in an extended theory for QM, what he hardly could accept. He often remarked: {\em `Itʼs a great puzzle to me ... behind the scenes something is going faster than the speed of light.'}
\end{quote}
For new perspectives on Bell's theorem and non-locality, the reader is referred to the recent papers by Brown and Timpson \cite{brown} and Zukowski and Brukner \cite{zuk1, zuk2}.
I will have occasion to return to this point later while discussing entanglement and the violation of Bell-like inequalities in classical optics. 

{\flushleft{\em The Kochen-Specker Theorem}}
\vskip 0.1in

Unlike Bell who considered locality as a criterion, Kochen and Specker considered non-contextual hidden variables, i.e. variables that depend only on the quantum system being measured and not on the measuring device \cite{ks}. This idea of non-contextuality is analogous to the classical notion that measurements only {\em reveal values of pre-existing properties} and do not depend on prior measurements of other compatible properties. Their theorem is based on two axioms: (i) {\em value definiteness} (i.e. all observables have definite values at all times) and (ii) {\em non-contextuality} (i.e. the value of an observable is independent of its measurement context). They showed that theories with such hidden variables were incompatible with quantum mechanics when the dimension of the Hilbert space is three or more. Together with the previous ones, this theorem ruled out all hidden variable theories except those involving contextual and non-local hidden variables such as the de Broglie-Bohm theory. The original proof of the theorem is fairly complex, but a simplified version for two spin $1/2$ particles has been given by Mermin \cite {mermin} and Peres \cite{peres}.

The violation of the Kochen-Specker theorem has ben demonstrated in neutron self-interference experiments \cite{hasegawa}.

{\flushleft{\em The Leggett-Garg Inequality}}
\vskip 0.1in
Leggett and Garg deduced a mathematical inequality that all macrorealistic physical theories must fulfil \cite{leg}. Macrorealistic theories are defined by the conjunction of two postulates:

(i) {\em Macroscopic realism}: A macroscopic object, which has available to it two or more macroscopically distinct states, is at any given time in a definite one of those states.
   
(ii) {\em Noninvasive Measureability}: It is possible in principle to determine which of these states the system is in without any effect on the state itself, or on the subsequent system dynamics.

Quantum mechanical systems obviously violate this inequality on both counts, and the original motivation for this work was to propose a test for quantum coherence in macroscopic systems. Unlike the Bell theorem which tests entanglement between spatially separated systems, this inequality tests the correlations in a single system at different times. Violations of this inequality therefore test either the impossibility of a realistic description of the system or the impossibility of measuring the system without disturbing it. Recently a number of experimental tests have been performed on a variety of microscopic systems such as superconducting qubits, nuclear spins, and photons and violations of these inequalities have been observed \cite{emary}. 

\section{Is the Wave Function Epistemic?}

The interpretation of a wave function in quantum mechanics, as we have seen. has been debated ever since its inception.  While some have argued for an {\em ontic} (state of reality) interpretation, others have preferred an {\em epistemic} (state of knowledge) interpretation. Of late quantum information processing theory seems to be favouring the latter interpretation \cite{qi, spekkens}. An advantage of an epistemic interpretation is that a sudden change or collapse of the wave function is seen only as a Bayesian updating on receipt of new information. There is therefore no measurement problem and non-locality in this interpretation.  

However, Pusey, Barrett and Rudolph \cite{pbr} have recently proved a no-go theorem with the help of a couple of reasonable assumptions (preparation independence) to rule out an epistemic interpretation. That was followed by Collbeck and Renner showing that if measurement settings can be chosen freely, a system's wave function is in one-to-one correspondence with its elements of reality, thus ruling out any epistemic or subjective interpretation \cite{cr}. However, Lewis {\em et al} \cite{pg} have shown that an epistemic interpretation is possible if one drops the preparation independence assumption and also slightly weakens the definition of an epistemic state given by Harrigan and Spekkens \cite{har}. Ghirardi and Romano \cite{ghirardi} have criticized the assumptions of free choice and the completeness of quantum mechanics made by Collbeck and Renner. The debate is therefore far from settled. The current situation has been reviewed by Leifer \cite{leifer}.

\section{Wave Functions in Classical Physics}
{\flushleft{\em Classical Mechanics}}
\vskip 0.1in
The fact that even Newtonian dynamics can be formulated in a Hilbert space spanned by square integrable functions but with a set of commuting hermitian operators as observables, has been known since the early 1930s, but only to a handful of physicists. I am referring to the classic works of Koopman \cite{K} and von Neumann \cite{vN} who developed a complete phase space theory of classical mechanics. The dynamics is Liouvillian. 

Though the classical Koopman-von Neumann (KvN) wave functions are complex, their relative phases are unobservable in classical mechanics. This is inherent in the way the KvN formalism is constructed. The basic idea of the Koopman theory \cite{K} lay in the classical phase space description of statistical mechanics.  For simplicity (but without loss of generality) let us consider a single-particle system described by a Hamiltonian $H(p,q)$ in 2-dimensional classical phase space wih commuting position and momentum variables $(\hat{q}, \hat{p})$ and the distribution function $f(q,p;t)$. The distribution obeys the classical Liouville equation
\beq
\frac{\partial f(q,p;t)}{\partial t} = \left(\frac{\partial H}{\partial q}\frac{\partial}{\partial p} - \frac{\partial H}{\partial p}\frac{\partial}{\partial q}\right)f(q,p;t) \equiv -i\hat{L}f(q,p;t) \label{1}
\eeq 				
This was followed by von Neumann \cite{vN} and Koopman postulating that this dynamics can be viewed as arising from underlying classical square integrable wave functions $\psi(q,p;t)$ in the Hilbert space of classical phase space variables obeying the equation										
\beq
\frac{\partial \psi(q,p;t)}{\partial t} = \left(\frac{\partial H}{\partial q}\frac{\partial}{\partial p} - \frac{\partial H}{\partial p}\frac{\partial}{\partial q}    \right)\psi(q,p;t)= -i\hat{L}\psi(q,p;t).\label{2}
\eeq 					
The complex conjugate wave function $\psi^*(q,p;t)$ is postulated to obey the equation
\beq
\frac{\partial \psi^*(q,p;t)}{\partial t} = \left(\frac{\partial H}{\partial q}\frac{\partial}{\partial p} - \frac{\partial H}{\partial p}\frac{\partial}{\partial q}    \right)\psi^*(q,p;t)=-i\hat{L}\psi^*(q,p;t).\label{3}
\eeq 
The scalar product in the Hilbert space is defined by
\beq
\langle \phi|\psi\rangle = \int dp\, dq\, \phi^*(q,p)\psi(q,p),
\eeq				
and the integrability condition is given by
\beq
||\psi||^2=\int_{-\infty}^{\infty}\int_{-\infty}^\infty dq\, dp\, \psi^*(q,p;t)\psi(q,p;t) = N.
\eeq 
Using dimensionless variables and identifing the density in phase space by
\beq
\psi^*(q,p;t)\psi(q,p;t)=|\psi(q,p;t)|^2 \equiv \rho(q,p;t),\label{rho}
\eeq
and using eqs.(\ref{2}) and (\ref{3}), one can easily verify that the phase space density obeys the equation 
\beq
\frac{\partial \rho(q,p;t)}{\partial t} = \left(\frac{\partial H}{\partial q}\frac{\partial}{\partial p} - \frac{\partial H}{\partial p}\frac{\partial}{\partial q}    \right)\rho(q,p;t) \Rightarrow i\frac{\partial \rho}{\partial t} = \hat{L}\rho\label{5}
\eeq
which is the Liouville equation. Thus, the Liouville equation can be derived from the postulated eqns. (\ref{2}) and (\ref{3}) for the classical wave function $\psi(q,p;t)$ and its complex conjugate, showing that the classical phase space probability density can be recovered from the dynamics of the underlying wave function $\psi$ and $\psi^*$. Since only the probability density $\rho(q,p;t)$ is observable in classical phase space, the phase of the wave function is not observable in this theory. This guarantees that no interference effects occur with classical particles even though a complex wave function is associated with it. This can be made clearer by writing $\psi=\sqrt{\rho}\, {\rm exp}(iS)$ with $S$ as the phase. Insering this expression in equations (\ref{2}) and (\ref{3}) and separating the real and imaginary parts, one gets
\beq
i\frac{\partial \sqrt{\rho}}{\partial t} = \hat{L}\sqrt{\rho},\,\,\,\,i\frac{\partial S}{\partial t} = \hat{L}S \label{L2},
\eeq
showing that the amplitude and phase evolve independently. This results in a `superselection rule', and it is possible to work with only $\sqrt{\rho}$.

Position and momentum can be represented by self-adjoint and commuting operators $\left[\hat{q},\hat{p}\right]=0$ which have continuous spectra from $-\infty$ to $+\infty$, 
\beq
\hat{q}|q,p\rangle = q|q,p\rangle,\,\,\,\,\hat{p}|q,p\rangle = p|q,p\rangle,\label{a1}
\eeq
where $\{|q,p\rangle\}$, the set of simultaneous eigenstates of $\hat{q}$ and $\hat{p}$, span the KvN Hilbert space, are orthonormal and form a complete set:
\ben
\langle q^{'},p^{'}|q^{''},p^{''}\rangle &=& \delta(q^{'} - q^{''})\delta(p^{'} - p^{''}),\\
\int \int dq\,dp\,|q,p\rangle\langle q,p|&=&1.
\een

Later Sudarshan \cite{sud} showed that it is also possible to view classical mechanics as a `hidden variable' quantum theory with only the positive square root $\sqrt{\rho}= \sqrt{|\psi|^2}$ associated with the wave function $\psi$, its phase being unobservable. A {\em superselection rule} operates in this theory, making transitions between wave functions with different phases unobservable. In addition to the commuting position and momentum operators $\hat{q}$ and $\hat{p}$, he introduced an additional pair of canonical operators $\hat{\chi} = i\partial_p$ and $\hat{\pi} = -i\partial_q$ (putting $\hbar =1$) in the Hilbert space with the properties
\ben
\left[\hat{q}, \hat{\chi}\right] &=& 0,\,\,\,\,\,\,\,\, \left[\hat{q},\hat{\pi}\right] = i,\\
\left[\hat{p},\hat{\pi}\right] &=& 0,\,\,\,\,\,\,\,\, \left[\hat{p}, \hat{\chi}\right] = -i. 
\een
The dynamical variables $\omega = (\hat{q},\hat{p})$ and $\pi = (-\hat{\pi}, \hat{\chi}) = i \partial_{\omega}$ may be considered as the canonical coordinates and momenta of a quantum system with two degrees of freedom. The equations of motion of the classical system can be viewed as the equations of motion of a quantum system with the Hamiltonian operator
\ben
\hat{H} &=& i\frac{\partial H(\omega)}{\partial \omega^\mu}\epsilon^{\mu\nu}\frac{\partial}{\partial \omega^\nu},\nonumber\\
\epsilon^{\mu\nu} (\omega) &=& \left[\omega^\mu, \omega^\nu \right]_{PB},\label{H}
\een
and have the form
\beq
\dot{\omega} = -i \left[\omega, \hat{H}(\omega) \right] \equiv i(\omega\, \hat{H} - \hat{H}\,\omega) \label{QE1}
\eeq
Since the Hamiltonian is linear in the quantum momenta $\pi = i\partial_\omega$, every phase space density $\rho(\omega)$ is mapped into a new phase space density $\tilde{\rho}(\omega)$ such that $\tilde{\rho}(\tilde{\omega}) = \rho(\omega)$ where $\tilde{\omega}$ are the displaced values obtained by solving Eqn. (\ref{QE1}). In the Heisenberg picture the result is 
\beq
f(\omega) \rightarrow f(\tilde{\omega}). \label{QE2} 
\eeq
Now, the quantum operators $\pi = i \partial_\omega$ are made {\em unobservable} {\em at all times and under all conditions} by imposing a {\em superselection rule} that renders only the commutative algebra of functions $f(\omega)$ of the commuting dynamical varaibles $\omega = (q,p)$ observable. This is a construction that embeds classical mechanics in a quantum theory with {\em hidden variables}. Since vectors in the Hilbert space of the quantum system are represented in the Schr\"{o}dinger picture by wave functions $\psi(\omega)$, and because the relative phases of the ideal eigenstates of the coordinate operators are not measureable and therefore irrelevant, one has the equivalence
\beq
\psi(\omega ) \sim  \psi(\omega )e^{i\phi(\omega)}.
\eeq
Hence, only the absolute value of $\psi(\omega)$ is relevant and may be taken as the positive square root of the phase space density
\beq
\psi(\omega) = \sqrt{\rho(\omega)}
\eeq
A classical state represented by a point in phase space is identified with an ideal joint eigenstate of the coordinate operators, and its time development is given by the solutions of equations (\ref{QE1}) and (\ref{QE2}) describing an {\em observable} trajectory in phase space. It must be pointed out that the Hamiltonian operator $\hat{H}$ (eqn. \ref{H}) is {\em not observable}. What is observable is the Hamiltonian function $H(\omega)$ which is a function of the observable coordinate operators $\omega$ only.

Since the overall phase of a state is not observable either in quantum mechanics or in classical optics, all states which differ only by a phase factor are usually identified to get `projective rays', resulting in the space $CP = H/U(1)$ of rays. The Bloch sphere in quantum mechanics and the Poincare sphere in classical polarization optics are examples. However, relative phases are observable in both quantum mechanics and classical optics, though not in classical mechanics. Hence, to rule out interference effects in classical mechanics, a further projection or identification is required, namely, all states that differ by relative phases are also identified. This results in the quotient space $CP^* = CP/U(1)$ which implements the required {\em superselection rule}. 

As an example of this method of treating classical mechanics as quantum mechanics with a hidden variable, let us consider the Hamiltonian operator of a free particle,
\beq
\hat{H}(\hat{q},\hat{p},\hat{\chi}, \hat{\pi}) = \frac{\hat{p}\hat{\pi}}{m}. \label{Hop} 
\eeq
The Heisenberg equations of motion are
\ben
\dot{\hat{p}} &=& i\left[\hat{H}, \hat{p}\right] =0,\nonumber\\
\dot{\hat{q}} &=& \frac{\hat{p}}{m},\nonumber\\
\dot{\hat{\pi}} &=& 0,\nonumber\\
\dot{\hat{\chi}} &=& i[\hat{H}, \hat{\chi}].
\een
The equations of motion for the observable dynamical variables $(\hat{q},\hat{p})$ are identical to Hamilton's equations, but the dynamical variables $(\hat{\chi}, \hat{\pi})$ are not observables. We will see later how this quantum mechanical treatment of a classical system can be used to deal with the case when it interacts with a quantum mechanical system.

The theory has been developed further since 1998 by a number of authors \cite{kvn1,kvn2,kvn3,kvn4,kvn5,mauro,kvn6,kvn7,kvn8,kvn9,kvn10,kvn11,kvn12,kvn13, kvn14}. They introduce the additional operators $\hat{\lambda}_q = -i\partial_q$ and $\hat{\lambda}_p = -i\partial_p$ which can be identified with the operators $\hat{\pi}$ and $-\hat{\chi}$ respectively. To go from the $(q,p)$ representation to the $(q,\lambda_p)$ representation with
\beq
\hat{q}|q,\lambda_p\rangle = q|q,\lambda_p\rangle,\,\, \hat{\lambda}_p|q,\lambda_p\rangle = \lambda_p|q,\lambda_p\rangle,
\eeq 
the required unitary transformation equations are \cite{mauro} 
\beq
-i\frac{\partial}{\partial p^{'}} \langle q^{'},p^{'}|q,\lambda_p\rangle = \lambda_p \langle q^{'},p^{'}|q,\lambda_p\rangle
\eeq
so that
\beq
\langle q^{'},p^{'}|q,\lambda_p\rangle = \frac{1}{\sqrt{2\pi}}\delta (q -q^{'})\,{\rm exp}\left(ip^{'}\lambda_p\right).\label{tr}
\eeq
Hence,
\beq
\langle q, \lambda_p|\psi\rangle = \int\int dq^{'} dp \langle q, \lambda_p|q^{'},p\rangle\langle q^{'},p|\psi\rangle =  \frac{1}{\sqrt{2\pi}} \int dp\, {\rm exp}\left(ip \lambda_p\right)\langle q,p|\psi\rangle,
\eeq
or equivalently,
\ben
\psi(q,\lambda_p,t) &=& \frac{1}{\sqrt{2\pi}} \int dp\, {\rm exp}\left(ip \lambda_p\right)\psi(q,p,t),\label{psi}.
\een
For a particle of unit mass in a potential $V$ the Liouville equation for the classical wave function in this representation is
\beq
i\frac{\partial}{\partial t}\psi(q,\lambda_p,t)=\left(\frac{\partial}{\partial q}\frac{\partial}{\partial \lambda_p} -V^{'}(q)\lambda_p \right)\psi(q,\lambda_p,t)\label{C1}
\eeq
and its complex conjugate is
\beq
i\frac{\partial}{\partial t}\psi^*(q,\lambda_p,t)=-\left(\frac{\partial}{\partial q}\frac{\partial}{\partial \lambda_p} -V^{'}(q)\lambda_p \right)\psi^*(q,\lambda_p,t).\label{C2}
\eeq
It follows from these equations that
\ben
\frac{\partial}{\partial t}\rho(q,\lambda_p,t) &=& -J(q,\lambda_p,t),\nonumber\\
J(q,\lambda_p,t) &=& i\left(\psi^*\frac{\partial}{\partial q}\frac{\partial}{\partial \lambda_p}\psi -\frac{\partial}{\partial q}\frac{\partial}{\partial \lambda_p}\psi^* \psi\right). \label{L3}
\een
Writing
\beq
\psi(q,\lambda_p,t)= \sqrt{\rho(q,\lambda_p,t)}\,{\rm exp}\,iS(q,\lambda_p,t),\label{L4}
\eeq
one sees that the time derivative of the density is connected with the wave function, unlike in the (q,p) representation (see eqn. \ref{L2}), and hence, phase features are present in this representation though not in the $(q,p)$ representation. However, {\em these phase features remain unobservable} because of the {\em superselection rule}.  

\vskip 0.1in
{\flushleft{\em Classical Electrodynamics}}
\vskip 0.1in
A fundamental difference between classical particles and classical waves (like electromagnetic waves) is that interference effects {\em are} observable in the latter. Hence, it should be possible to develop a phase space theory of such fields by using a projective Hilbert space $CP$ so that no superselection rule operates, and this has been done by Rajagopal and Ghose \cite{raja}. The classical `wave function' of the electromagnetic field can be written as a six-component column vector
\beq
\psi(E_i,B_i)= \left(\begin{array}{c}
E_x \\E_y\\E_z\\-B_x\\-B_y\\-B_z
\end{array} \right)  
\eeq
and its dual $\psi^\dagger (E_i,B_i) = \left(E^*_x,E^*_y, E ^*_z, -B^*_x, -B^*_y, -B^*_z\right)$. Denoting by $\varphi$ the complex dynamical variables $\{B_i, E_i\}$, one can define inner products
\beq
\langle \phi|\psi\rangle = \int \Pi_i dE_i\Pi_j dB_j \delta(\vec{E}.\vec{B})\langle \phi|E_i,B_i\rangle\langle E_i,B_i|\psi\rangle =  \int d\varphi\, \phi^\dagger(\varphi)\psi(\varphi)
\eeq 
where $d\varphi=\Pi_i dE_i \Pi_j dB_j\delta(\vec{E}.\vec{B})$ and the scalar product is given by 
\beq
||\psi||^2=\langle \psi|\psi\rangle = \int d\varphi\, \psi^\dagger (\varphi) \psi(\varphi) = \int d\varphi\,\rho(\varphi)
\eeq
with $\rho(\varphi) =\psi^\dagger (\varphi) \psi(\varphi)= \sum_i(E_i^*E_i + B_i^*B_i)$ ($i=x,y,z$). Normalized by the total energy, this gives the probability density in phase space if one treats the magnetic fields $\{B_i\}$ as the cordinates and the electric fields $\{E_i\}$ as the conjugate momenta such that $\vec{E}.\vec{B}=0$.  
Using units in which $\epsilon_0 = \mu_0 = c =1$, let the equations of motion for the wave functions be
\ben
\frac{\partial}{\partial t}\psi(\varphi,t) &=& -i\hat{L}\psi(\varphi,t) = -\sum_i\beta_i \partial_i\psi(\varphi,t),\label{eq1}\\
\frac{\partial}{\partial t}\psi(\varphi,t)^\dagger &=& -i\psi(\varphi,t)^\dagger\hat{L} =-\sum_i\partial_i\psi(\varphi,t)^\dagger\beta_i,\label{eq2} 
\een
where $\partial_i \equiv \partial/\partial {x}_i$ and $\hat{L}=-i\sum_i\beta_i \partial_i$ is the Liouvillian operator and $\beta_i^\dagger = \beta_i$ are three $6\times 6$ hermitian matrices (whose explicit forms are given in the Appendix) such that the equations (\ref{eq1}) and (\ref{eq2}) encode the Maxwell equations
\beq
\dot{\vec{E}} = {\rm curl}\,\vec{B},\,\,\,\,\dot{\vec{B}} = -{\rm curl}\,\vec{E}.
\eeq
Multiplying the first equation by $\psi(\varphi,t)^\dagger$ from the left and the second equation by $\psi(\varphi,t)$ from the right and adding them, one gets the continuity equation
\ben
\frac{\partial \rho(\varphi,t)}{\partial t} &=& - 2i\psi(\varphi,t)^\dagger\hat{L}\psi(\varphi,t) = - \sum_i\partial_i S_i(\varphi,t),\nonumber\\
S_i(\varphi,t) &=& \psi^\dagger(\varphi,t) \beta_i \psi(\varphi,t).\label{cont}
\een
Given the forms of the $\beta$ matrices $S_i=(\vec{E}\times \vec{B})_i$ turn out to be components of the Poynting vector. This is the Liouville equation for the phase density which is different from the equations of motion of the underlying wave functions. It is clear from this that the time derivative of the phase space density and amplitude $\psi$ are coupled, and the relative phase is an observable unlike in the KvNS theory of classical mechanics.

As in classical mechanics, it is possible to embed classical electrodynamics also in a quantum field theory by doubling the number of variables, the additional variables being {\em hidden variables}. Let $\hat{\chi}_i \equiv -\hat{\lambda}_{E_i} = i\partial_{E_i}$, $\hat{\chi}^*_i \equiv -\hat{\lambda}_{E^*_i} = i\partial_{E^*_i}$,  $\hat{\pi}_i \equiv \hat{\lambda}_{B_i} = -i\partial_{B_i}$ and $\hat{\pi}^*_i \equiv \hat{\lambda}_{B^*_i} = -i\partial_{B^*_i}$ ($\partial$ now denoting functional partial derivatives) be these hidden canonical variables with the commutation properties
\ben
\left[\hat{B}_i(x), \hat{\chi}_j(x^\prime)\right]_{t=t^\prime} &=& 0,\,\,\,\,\,\,\,\, \left[\hat{B}_i(x),\hat{\pi}_j(x^\prime)\right]_{t=t^\prime} = i \delta^3(x - x^\prime),\nonumber\\
\left[\hat{E}_i(x),\hat{\pi}_j(x^\prime)\right]_{t=t^\prime} &=& 0,\,\,\,\,\,\,\,\, \left[\hat{E}_i(x), \hat{\chi}_k(x^\prime)\right]_{t=t^\prime} = -i \delta^3(x - x^\prime)\nonumber,\\
\left[\hat{B}^*_i(x), \hat{\chi}^*_j(x^\prime)\right]_{t=t^\prime} &=& 0,\,\,\,\,\,\,\,\, \left[\hat{B}^*_i(x),\hat{\pi}^*_j(x^\prime)\right]_{t=t^\prime} = i \delta^3(x - x^\prime),\nonumber\\
\left[\hat{E}^*_i(x),\hat{\pi}^*_j(x^\prime)\right]_{t=t^\prime} &=& 0,\,\,\,\,\,\,\,\, \left[\hat{E}^*_i(x), \hat{\chi}^*_k(x^\prime)\right]_{t=t^\prime} = -i \delta^3(x - x^\prime)\label{A}. 
\een  
The conventional classical representation corresponds to the choice $\hat{\varphi}=(\hat{B},\hat{E})$ as the complete set of observable dynamical variables with $\hat{B}_i$ as the cordinates and $\hat{E}_i$ as the momenta and 
\beq
\left[\hat{E}_i, \hat{B}_j\right] =0,\,\,\,\,\,\, \left[\hat{E}^*_i, \hat{B}^*_j\right] =0\,\,\,\,\,\,\,\,\forall i,j=1,2,3. 
\eeq
Other quantized representations can also be used by choosing a pair of commuting variables from $\hat{\varphi}$ and $\hat{\Pi} = (\hat{\chi}, \hat{\pi}) = i \partial_{\varphi}$ as the coordinates and momenta. For example, one can make a unitary transformation from the commuting variables $\hat{\varphi}=(\hat{B},\hat{E})$ to the commuting variables $\hat{\xi}= (\hat{B}, \hat{\lambda}_E= -\hat{\chi})$ with the properties
\beq
\hat{B_i}|B,\lambda_E\rangle = B_i|B,\lambda_E\rangle,\,\, \hat{\lambda}_{E_i}|B,\lambda_E\rangle = \lambda_{E_i}|B,\lambda_E\rangle
\eeq
and the transformation equation 
\beq
\langle B^{'},E^{'}|B,\lambda_E\rangle = \frac{1}{(2\pi)^{3/2}}\Pi_i\delta (B_i -B_i^{'})\,{\rm exp}\left(i\sum_iE_i^{'}\lambda_{E_i}\right).\label{tr2}
\eeq
and
\ben
\psi(\xi,t) &=& \frac{1}{(2\pi)^{3/2}} \int d^3 E\, {\rm exp}\left(i\sum_iE_i \lambda_{E_i}\right)\psi(\varphi,t).\label{chi}
\een 
The Liouville equations for this classical wave function and its adjoint are 
\ben
i\frac{\partial }{\partial t}\psi(\xi,t) &=& \left(\frac{\partial}{\partial B_i^*}\frac{\partial}{\partial \lambda_{E_i}} - B_i^*\lambda_{E_i} \right)\psi(\xi,t),\label{CEM1}\\
i\frac{\partial }{\partial t}\psi^*(\xi,t) &=& -\left(\frac{\partial}{\partial B_i^*}\frac{\partial}{\partial \lambda_{E_i}} - B_i^*\lambda_{E_i} \right)\psi^*(\xi,t).\label{CEM2}
\een
Defining $\rho(\xi,t) = \psi^*(\xi,t)\psi(\xi,t)$, it follows that
\ben
\frac{\partial }{\partial t}\rho(\xi,t) &=& -J(\xi,t),\\
J(\xi,t) &=& i\left(\psi^*\frac{\partial}{\partial B_i^*}\frac{\partial}{\partial \lambda_{E_i}}\psi -\frac{\partial}{\partial B_i^*}\frac{\partial}{\partial 
\lambda_{E_i}}\psi^* \psi\right).
\een
If one writes $\psi(\xi,t)$ in the form
\beq
\psi(\xi,t)= \sqrt{\rho(\xi,t)}\,{\rm exp}\,iS(B_i^*,\lambda_{E_i},t),
\eeq
it becomes clear that the time derivatives of the density and amplitude get connected, showing that phase features are preserved in the $(\hat{B}, \hat{\lambda}_{E})$ representation of the electromagnetic field. Alternatively, one could choose $(\hat{\lambda}_B =\hat{\pi}, \hat{E})$ as the commuting set. Both of these representations are examples of embedding classical electrodynamics in a quantum field theory with hidden variables. The difference from classical mechanics is that there is no {\em superselection rule} in electrodynamics making relative phases between different states unobservable, and hence superpositions of states can be written in electrodynamics.

In the full operator formalism the equations of motion in the Schr\"{o}dinger picture takes the form
\ben
\frac{\partial}{\partial t}\hat{\psi}(\hat{\varphi},t) &=& -\sum_i\beta_i \partial_i\hat{\psi}(\hat{\varphi},t),\label{eq1a}
\een
which are the Maxwell equations
\beq
\dot{\vec{\hat{E}}} = {\rm curl}\,\vec{\hat{B}},\,\,\,\,\dot{\vec{\hat{B}}} = -{\rm curl}\,\vec{\hat{E}}.
\eeq
in the operator form, and
\ben
\frac{\partial}{\partial t}\hat{\psi}(\hat{\Pi},t) &=& -\sum_i\beta_i \partial_i\hat{\psi}(\hat{\Pi},t),
\een
which are the equations of motion for the hidden variables $\hat{\Pi}= (\hat{\chi},\hat{\pi})$:
\beq
\dot{\vec{\hat{\chi}}} = {\rm curl}\,\vec{\hat{\pi}},\,\,\,\,\dot{\vec{\hat{\pi}}} = -{\rm curl}\,\vec{\hat{\chi}}.
\eeq

An extremely important development in this field is the interpretation of the Wigner function as a KvNS wave function, i.e. as a probability amplitude that need not be positive everywhere \cite{bondar, bond2}. The significance of the Wigner function so far has been that it mimics the classical distribution function, which is positive everywhere, but with characteristic quantum features imbedded in it, namely that it is not positive everywhere, the quantumness being conventionally attributed to its negative features. Bondar and his coworkers have given it a new interpretation as a wave function and hence as a `a probability amplitude for the quantum particle to be at a certain point of the classical phase space'. This dissolves the `mystery' of the negativity of the Wigner function. According to this interpretation the essential quantumness of a process lies not in the negativity of the Wigner function but in the distinctiveness of quantum processes to make transitions from positivity to nonpositivity and vice versa while classical processes are only negativity and positivity preserving. Hence, negativity and positivity preserving quantum processes can be simulated by classical polarization optics. 

\section{Entanglement and Non-locality}

Entanglement and Bell violations have hitherto been exclusively associated with quantum phenomena and non-locality, and so it might at first sight appear absurd to claim that they occur in the very bastion of classical physics, namely classical optics, and that they do not imply non-locality. This is understandable because generations of physicists have been brought up on the myth that Hilbert spaces occur only in quantum physics, and that violations of Bell inequalities occur only in quantum phenomena and imply non-locality. The fact is that Hilbert spaces and the Schmidt decomposition were known in classical physics before quantum mechanics was even born \cite{schmidt}. Furthermore, and this is a subtle point that needs to be emphasized, apart from realism, {\em `separability' of states in different Hilbert spaces is sufficient for CHSH-type inequalities}, and it is not necessary to assume locality, as Bell did. Einstein was aware of the difference between separability and locality, and that is clear from his `Autobiographical Notes' \cite{ein}. He considers two spatially separated and non-interacting partial systems $S_1$ and $S_2$ described by an entangled wave function $\psi$. When a measurement is made in either of two mutually incompatible bases on $S_1$, two different wave functions for $S_2$ result, as we have seen in Section 2. According to Einstein,
\begin{quote}
Now it appears to me that one may speak of the real factual situation of the partial system $S_2$. Of this real factual situation, we know to begin with, before the measurement of $S_1$, even less than we know of a system described by the $\psi$-function. But, on one supposition we should, in my opinion, absolutely hold fast: the real factual situation of the system $S_2$ is independent of what is done with the system $S_1$, which is spatially separated from the former. 
\end{quote}
Since a unique wave function cannot be associated with the `real factual situation' of $S_2$, the quantum mechanical description of the system is incomplete. Einstein then continues to say:
\begin{quote}
One can escape from this conclusion only by either assuming that the measurement of $S_1$ (telepathetically) changes the real factual situation of $S_2$ or by denying independent real situations as such to things which are spatially separated from each other. Both alternatives appear to me entirely unaceeptable.
\end{quote}
Einstein is clearly setting out the {\em ontic} view of the systems and their {\em unique} quantum mechanical wave functions to show that such a view leads to an incomplete description of reality unless one admits either telepathy {\em or} non-separability. He clearly distinguishes between `telepathy' (i.e. action-at-a-distance between {\em independent} and {\em non-interacting} systems) and `denial of independence' to even spatially separated non-interacting systems (non-separability), something he could not do in the single-particle case (see Section 2). He, however, rejects both. In his time nobody was aware that the spatial and polarization modes of a {\em single} beam of classical light, which belong to different and independent Hilbert spaces, could become `non-separable'. There is obviously no question of any `non-locality' in such cases. Hence, we can say today, thanks to recent developments in classical optics, that {\em Bell-CHSH-like violations are not sure tests of either quantumness or non-locality.} What the Bell theorem does show is that only non-separable states violate Bell-like inequalities, not separable or product states, and that too for very special apparatus settings only. This is true in both classical polarization optics and quantum physics.

There is, however, one clear difference between classical optics and quantum mechanics coupled with the projection postulate (von Neumann's `process 1') which would require that a strong projective measurement on the partial system $S_1$ in either basis would result in one of the two terms on the right-hand sides of Eqn. (\ref{EPR}) {\em disappearing instantly} in every measurement. This does not happen in either classical optics or in interpretations of quantum mechanics that do not invoke von Neumann's `process 1' (or collapse) such as the Everett interpretation. Clearly therefore a real factual change of the ontic state of $S_2$ can be inferred only provided a strong projective measurement does occur as a result of a measurement on a distant non-interacting partial system $S_1$. This putative process is precisely the one that many interpretations of quantum mechanics avoid, as we have seen. Alternatively, one must admit that two (or more) wave functions can describe the same ontic state of the system $S_2$. The first alternative implies non-locality whereas the second alternative implies that quantum mechanical wave functions are {\em epistemic} (i.e. states of knowledge) and provide an incomplete description of nature. Einstein was in favour of the latter view \cite{har}. As we have seen, there is much current interest in this area of quantum physics \cite{leifer}. Classical KvNS wave functions can be ontic in spite of CHSH-like violations because there is no collapse in classical measurements.

\section{Realism and Non-contextuality}

What about realism? Classical realism is associated with non-contextuality, i.e. the notion that a physical property is independent of the context in which it is measured. In other words, a measurement result is predetermined and is not affected by whether previous or simultaneous measurement of any other compatible or co-measureable observable is carried out on the system or not. In classical physics all observables are, of course, compatible. This is why local realistic classical theories have been presumed to be non-contextual because the result of a measurement in such theories does not depend on any measurement made simultaneously on a spatially separated (mutually non-interacting) system. This concept can be tested by joint measurements of compatible observables that are not necessarily spatially separated such as the spatial and polarization modes of a single beam of  classical light---a change of the polarization (spatial) mode of a separable light beam should not change or affect its spatial (polarization) mode, showing that such a beam is non-contextual, but this should not be true of an entangled beam. It should be fairly easy to test this by using Laguerre-Gaussian modes. It should also be possible to test the violation of a CHSH-type inequality that can be deduced by assuming separability, as well as the Kochen-Spekker theorem \cite{ghose2}.

This renders the classical concept of reality problematic---one has to admit that reality can be contextual in certain cases even in classical physics! 

\section{Waves, Particles and Quantum Measurement}
We have seen that complex wave functions alone do not distinguish quantum physics from classical physics. The fundamental difference lies in the commutation properties of the observable dynamical variables and the dynamical evolution equations. In classical physics the evolution equation is the Liouville equation
\beq
\frac{\partial \rho_c}{\partial t} = -i\hat{L}\rho_c \label{L}
\eeq
where $\rho_c$ is the classical density in phase space and $\hat{L}$ is the Liouville operator. In quantum physics the corresponding evolution equation is the Moyal equation for the Wigner function,
\beq
\frac{\partial W(q,p,t)}{\partial t} = - \{\{W(q,p,t), H(q,p)\}\} 
\eeq
where $W(q,p,t)$ is the Wigner function, $H(q,p)$ is the Hamiltonian and $\{\{W(q,p,t), H(q,p)\}\}$ is their Moyal bracket. In the limit $\hbar \rightarrow 0$ the Moyal bracket reduces to the Poisson bracket, and the Moyal equation for the Wigner function reduces to the classical Liouville equation (\ref{L}). We have already seen (in the last paragraph of Section 6) that the Wigner function is, in fact, a probability amplitude for a quantum particle to be at a certain point of the classical phase space, and that in the classical limit it transforms into a classical KvNS wave function rather than into a classical probability distribution $\rho_c$.
The reader would recall that in the $(q,p)$ representation of the KvNS theory, the wave function, in fact, obeys the same Liouville equation as the density $\rho_c$ (Eqns (\ref{1}), (\ref{2}) and (\ref{3})) and its phase is unobservable. 

This brings us back to wave-particle duality which is widely believed to be the most typical example of a quantum paradox. It is not so well known that classical physics also has a conceptually dual character, as Einstein pointed out in his `Autobiographical Notes' \cite{ein} in which he writes,
\begin{quote}
$\cdots$ there exist two types of conceptual elements, on the one hand, material points with forces at a distance between them, and, on the other hand,  the continuous field. It represents an intermediate state in physics without a uniform basis for the entirety, which---although unsatisfactory---is far from having been superseded.
\end{quote} 
It is widely held that this dual character is resolved in quantum field theory in which `particles' appear as quantized excitations of a continous field. 

It should be clear from the above discussion that the KvNS theory of classical physics already provides the underlying unity of waves and particles that Einstein referred to as the `uniform basis for the entirety'. We have seen in Section 6 that a {\em superselection rule} operates in the KvNS theory which renders the relative phases between complex classical wave functions unobservable, ensuring particle-like trajectories in phase space, and that the corresponding Hilbert space is $CP^*=CP/U(1)$. Such a superselection rule, however, does not operate in the $CP$ Hilbert space spanned by the wave functions of the classical electromagnetic field. Leaving aside the case of the electromagnetic field (whose quantized version is quantum field theory), one is entitled to ask: what happens in the non-relativistic quantum mechanical case where the relative phases of the wave function are observable? There is no particle aspect of such a wave function until {\em a detection occurs} when the wave function appears to collapse into a single point, a `particle'. 

In quantum mechanics the dynamical variables form a non-commuting algebra so that not all dynamical variables can be measured simultaneously with precision. However, every measurement selects a commuting subalgebra of dynamical variables which are compatible and can be measured simultaneously with precision. So, the measurement process is such that it produces classical pointer readings and unambiguous measurements of a compatible set of dynamical variables. Hence, as Bohr emphasized, {\em quantum theory presupposes classical systems} which can be influenced by quantum systems. This is possible if the quantum system and the classical system comprising the apparatus can be coupled. However, as we have seen, the dynamical equations for classical and quantum systems are different. It is precisely this point, largely glossed over, that is addressed by Sudarshan's approach \cite{sud, sud2, sud3, sud4, sud5}. As we have seen, he proposed that the KvNS formalism can be looked upon as an embedding of the classical system in a quantum system with a continuum of superselection sectors. In other words, it is possible to look upon {\em classical mechanics as quantum mechanics with essentially hidden variables}. These are not the usual dynamical variables, but they are canonical, and hence they can be considered as dynamical variables. The equation of motion for the set of dynamical variables is given by equation (\ref{QE1}) which has the same form as the quantum mechanical equation of motion. This makes it possible to couple a quantum system to a classical system consistently.

Now consider a quantum mechanical harmonic oscillator with the Hamiltonian
\beq
\hat{H}_0 = \left[\frac{\hat{p}^2}{2m} + m\omega^2\frac{\hat{q}^2}{2}\right]
\eeq 
and the measurement of its momentum $p$ by the coupling term $g\hat{p}\hat{\chi}^A$ to a classical measuring apparatus $A$ with the Hamiltonian operator
\beq
\hat{H}^A = \frac{\hat{q}^A \hat{\chi}^A}{M},
\eeq 
which is a quantum mechanical system with hidden variables and a superselection rule (see equation (\ref{Hop})). Then the total Hamiltonian is
\beq
\hat{H} = \hat{H}_0 + \hat{H}^A + g\hat{p}\hat{\chi}^A,
\eeq
and the equations of motion are
\ben
\dot{\hat{q}} &=& \frac{\hat{p}}{m} + g\hat{\chi}^A,\\
\dot{\hat{p}} &=& m\omega^2 \hat{q},\\
\dot{\hat{q}}^A &=& 0, \\
\dot{\hat{p}}^A &=& \frac{\hat{\chi}^A}{M} + g\hat{p}.
\een
The position of the oscillator cannot be measured because the Hamiltonian $\hat{H}$ and $\dot{\hat{q}}$ contain the unobservable $\hat{\chi}^A$. This is consistent with Bohr's position that that there is an {\em uncontrollable disturbance} in the conjugate variable $\hat{q}$ when $\hat{p}$ is being measured. On the other hand, the equation of motion for $\hat{p}^A$ contains a term proportional to $\hat{p}$ and is an observable. Sudarshan has shown generally that the theory naturally leads to a measurement {\em via} the trajectory of {\em only a commuting set of quantum dynamical variables}.

Another example is the Stern-Gerlach experiment to measure the spin of a neutral particle. In this case the Hamiltonian operator is
\ben
H &=& -\frac{i}{m}\vec{p}.\vec{q} - i\Gamma S_3\frac{\partial}{\partial p_3} - \gamma B_3 S_3
\een
where $\Gamma$ is proportional to the magnetic field gradient. The equations of motion can be soved to give
\ben
q_1(t) &=& \frac{1}{m} p_1 t, \,\,\,\,\,\,\,\,q_2(t) = 0,\\
q_3(t) &=& \pm \frac{1}{2}\Gamma t^2
\een
with the spin taken as $\pm \frac{1}{2}$. This gives two parabolas for the trajectory.

As we have seen earlier, the superselection rule implies that one cannot write linear superpositions of classical KvNS states, and that their density matrix is mixed. The result of the coupling between the quantum system and the classical measuring apparatus is obviously therefore a mixed state
\beq
\hat{\rho} = \sum_i |c_i|^2 |\Phi_i\rangle\langle \Phi_i|
\eeq
where $|\Phi_i\rangle = |q\rangle_i\otimes|\omega\rangle_i$ with $|q\rangle_i$ a quantum state and $|\omega\rangle_i$ its correlated classical state. This is, indeed, the result that one would obtain by using quantum wave functions for the measuring apparatus and invoking von Numann's `process 1' (the collapse postulate). Hence, the KvNS theory for the classical apparatus is sufficient to lead to the standard quantum mechanical results without requiring collapse of the quantum state.

The superposition principle is fundamental to quantum mechanics and is a consequence of its linearity. It is reponsible for interference phenomena, famously exemplified by the double-slit experiment. These interference effects can in certain situations get suppressed (sometimes spontaneously), and the quantum systems are said to decohere. The theory of decoherence is the study of interactions between a quantum system and its environment (usually modelled by a heat bath) that lead to such suppression of interference \cite{zurek, zurek2}. The coupling of a measuring apparatus (treated quantum mechanically) to its environment would accordingly cause its rapid decoherence, resulting in a mixed state. 

Decoherence is usually regarded as being relevant to a variety of questions ranging from the measurement problem to irreversibility, and most importantly to the emergence of the classical world from a quantum substratum \cite{zurek3, zurek4}. It was originally claimed that decoherence also solves the measurement problem, but this claim has been refuted \cite{zeh, adler}. The reader may find Ref. \cite{bac2} a useful review article on decoherence.

Mixed states in the KvNS theory of classical mechanics are analogous to mixed states after decoherence in quantum mechanics, but have their origin in `hidden dynamical variables' or unobservable relative phases resulting from a superselection rule. Sudarshan's use of this (so far largely ignored) mechanism for decoherence of classical mechanical systems in solving the quantum measurement problem brings about a new harmony of waves and particles in quantum mechanics. 

\section{Concluding Remarks}
We began with a very brief review of the principal paradoxes of quantum theory leading to various interpretations of quantum mechanics as well as to the principal no-go theorems that have ruled out all hidden variable theories except those that are contextual and non-local such as the de Broglie-Bohm theory. We have seen how the nature of the quantum wave function, i.e. whether it is ontic or epistemic, has been the subject of a long debate which still continues. 

We have also seen how the concepts of waves and particles with all their subtleties and dichotomies have evolved since the inception of quantum theory, and also the principal attempts to harmonize them. Bohr viewed them as mutually exclusive but complementary aspects of a quantum entity whereas Einstein, de Broglie and Bohm preferred a more inclusive harmony. Some single photon experiments like the double prism experiment as well as the observation of predicted single photon trajectories in a double slit interference experiment using weak measurements lend considerable support to, though not conclusive proofs of, the latter view. It was Koopman and von Neumann who realized that complex wave functions spanning Hilbert spaces are not unique to quantum mechanics. They formulated classical mechanics in a manner analogous to quantum mechanics by using square integrable complex wave functions underlying the distribution function $\rho(q,p)$ in classical phase space of particle trajectories, and representing the associated physical observables by means of commuting self-adjoint operators, thus providing a clear and beautiful harmony of waves and particles. In their formulation, a {\em superselection rule} operates to make the relative phases of the wave functions unobservable. Sudarshan then showed how classical mechanics can be considered as a quantum theory with hidden variables. This necessitates a doubling of the dynamical variables with only two of them observable and the other two {\em hidden}. A useful way of looking at the {\em superselection rule} is to view it as a quotient map $CP \rightarrow CP/U(1)$. We have also seen how the KvNS theory can be extended to classical electrodynamics in a $CP$ Hilbert space in which relative phases are observable. This provides a sound foundation to classical electrodynamics that satisfactorily accounts for the entanglement/non-separability and CHSH-like violations already observed in classical polarization optics. Although the classical KvNS wave function is certainly ontic, surprisingly it can be contextual in certain cases, rendering the concept of classical realism problematic.

One of the important new insights that have been obtained through these developments is that {\em entanglement and CHSH-like violations are neither unique signatures of quantumness nor of non-locality}---they only signify an underlying Hilbert space structure and non-separability. There is a significant distinction between non-locality and non-separability, and what CHSH-like violations have shown is the latter, not necessarily the former. To establish non-locality in quantum physics, one must first establish the ontic character of collapse, which has not been done as yet. 

Another extremely important insight is the interpretation of the Wigner function as a KvNS wave function, i.e. a probability amplitude which need not be positive everywhere \cite{bondar, bond2}. This has important implications for simulating certain types of quantum information processes by using classical polarization optics.

The KvNS theory seems to suggest that classical physics emerges from a quantum substratum when the non-commuting dynamical variables somehow become {\em hidden}.  

\section{Acknowledgement}
I am most grateful to A. K. Rajagopal for reading several drafts of this paper and suggesting many improvements. However, all shortcomings that remain are entirely mine. I also thank the National Academy of Sciences, India for the grant of a Fellowship that enabled this work to be undertaken.

\vskip 0.1in
{\flushleft{\em Note added}}
\vskip 0.1in
My attention has just been drawn to the paper arxiv: 0501.03202 [quant-ph] by Jennings and Leifer dated 13th January, 2015. There is no reference to the developments in classical polarization optics related to entanglement and Bell-CHSH-like violations (Refs. 10-22) in this paper.

\section{Appendix}
The matrices $\beta_i$ are given by

$\beta_x =\left(\begin{matrix}
0&0&0&0&0&0\\0&0&0&0&0&-1\\0&0&0&0&1&0\\
0&0&0&0&0&0\\0&0&1&0&0&0\\0&-1&0&0&0&0
\end{matrix}\right)$,\,\,\,$\beta_y =\left(\begin{matrix}
0&0&0&0&0&1\\0&0&0&0&0&0\\0&0&0&-1&0&0\\
0&0&-1&0&0&0\\0&0&0&0&0&0\\1&0&0&0&0&0
\end{matrix}\right)$,\\
$\beta_z =\left(\begin{matrix}
0&0&0&0&-1&0\\0&0&0&1&0&0\\0&0&0&0&0&0\\
0&1&0&0&0&0\\-1&0&0&0&0&0\\0&0&0&0&0&0
\end{matrix}\right)$


\begin{thebibliography}{0}
\bibitem{gen}
M. Genovese, {\em Adv. Sc. Lett.} {\bf 3}, 249-258 (2010).
\bibitem{K}
B. O. Koopman,  {\em Proc. Natl. Acad. Sci. U.S.A.} {\bf 17}, 315 (1931).
\bibitem{vN}
J. von Neumann,  {\em Ann. Math.} {\bf 33}, 587 (1932); {\em ibid.} {\bf 33}, 789-791 (1932).
\bibitem{sud}
E. C. G. Sudarshan, {\em Pramana} {\bf 6}, 117-126 (1976).
\bibitem{sp}
R. J. C. Spreeuw, {\em Found. of Phys.} {\bf 28} 361 (1998); {\em Phys. Rev. A} {\bf 63}, 062302 (2001).
\bibitem{g}
P. Ghose \& M. K. Samal, arXiv:quant-ph/0111119v1 22Nov 2001 and references therein.
\bibitem{radial}
S. C. Tidwell, G. H. Kim \& W. D. Kimura, {\em Applied Phys. Lett.} {\bf 32}, 5222-5229 (1993).
\bibitem{oron}
R. Oron, S. Blit, N. Davidson \& A. A. Friesem, {\em Applied Phys. Lett.} {\bf 77}, 3322-3324 (2000).
\bibitem{kozawa}
Y. Kozawa \& S. Sato, {\em Optics Lett.} {\bf 30}, 3063-3065 (2005).
\bibitem{souza}
C. E. R. Souza, J. A. O. Huguenin, P. Milman, and A. Z. Khoury, {\em Phys. Rev. Lett.} {\bf 99}, 160401 (2007).
\bibitem{holleczek}
A. Holleczek, A. Aiello1, C. Gabriel1, C. Marquardt, G. Leuchs, arXiv:1012.4578v1 [quant-ph] 21 Dec 2010. 
\bibitem{borges}
C. V. S. Borges, M. Hor-Meyll, J. A. O. Huguenin \& A. Z. Khoury, {\em Phys. Rev. A} {\bf 82}, 033833 (2010). 
\bibitem{simon}
B. N. Simon, S. Simon, F. Gori, M. Santarsiero, R. Borghi, N. Mukunda \& R. Simon, {\em Phys. Rev. Lett.} {\bf 104}, 023901 (2010).
\bibitem{gabriel}
C. Gabriel, A. Aiello, W. Zhong, T. G. Euser, N. Y. Joly, P. Banzer, M. Fortsch, D. Elser, U. L. Andersen, C. Marquardt, P. S. J. Russell \& G. Leuchs, {\em Phys. Rev. Lett.} {\bf 106}, 060502 (2011). 
\bibitem{eberly}
Xiao-Feng Qian \& J. H. Eberly, {\em Optics Lett.} {\bf 36}, 4110-4112, (2011).
\bibitem{agarwal}
P. Chowdhury, A. S. Majumdar \& G. S. Agarwal, {\em Phys. Rev. A} {\bf 88}, 013830 (2013).
\bibitem{kagalwala}
K. H. Kagalwala, G. Di Giuseppe, A. F. Abouraddy and B. E. A. Saleh, {\em Nature Photonics} {\bf 7}, 72-78 (2013). 
\bibitem{ghose2}
P. Ghose \& A. Mukherjee, {\em Adv. Sc., Eng. and Med.},  {\bf 6}, 246-251 (2014).
\bibitem{pereira}
L. J. Pereira, A. Z. Khoury and K. Dechoum, {\em Quantum and classical separability of spin-orbit laser modes}, arxiv: 1409.0889, Sept 2014.
\bibitem{ghose}
Partha Ghose and  A. Mukherjee, {\em Rev. in Theoret. Sc.} {\bf 2}, 1-14 (2014); arXiv: 1308.6154.
\bibitem{aiello}
A. Aiello, F. T\"{o}ppel, C. Marquardt, E. Giacobino, and G. Leuchs, {\em Classical entanglement: Oxymoron or resource?}, arxiv: 1409.0213v2 [quant-ph] Dec 2014.
\bibitem{raja}
A. K. Rajagopal and Partha Ghose, {\em Hilbert Space Theory of Classical Electrodynamics}, arXiv: 1409.5874 [quant-ph] Sep 2014.
\bibitem{bondar}
D. I. Bondar, R. Cabrera, D. V. Zhdanov and H. A. Rabitz, {\em Phys. Rev.A} {\bf 88}, 052108 (2013); arXiv:quant.phys/1202.3628 (2012).
\bibitem{bond2}
D. I. Bondar, R. Cabrera, R. R. Lompay, M. Yu. Ivanov and H. A. Rabitz, {\em Phys. Rev. Lett.} {\bf 109},190403 (2012): arXiv:quant.phys/1105.4014 (2012).
\bibitem{feynman}
R. P. Feynman, R. B. Leighton, and M. Sands, {\em Feymnan Lectures on Physics}, Vol. III (Addison Wesley, Reading, Massachusetts, 1964), Chaper 1, section 6.
\bibitem{bohr}
N. Bohr, `Discussion with Einstein on Epistemological Problems in Atomic Physics' in {\em Albert Einstein: Philosopher-Scientist}, ed. P. A. Schilpp, pp. 200-41, The Library of Living Philosophers, Princeton (1949).
\bibitem{rauch1}
H. Rauch, W. Trimer and U. Bonse, {\em Phys. Lett. A} {\bf 47}, 369 (1974).
\bibitem{rauch2}
H. Rauch and J. Summhammer, {\em Phys. Lett. A} {\bf 104}, 44 (1984).
\bibitem{rauch3}
G. Badurek, H. Rauch and D. Tuppinger, {\em Phys. Rev A} {\bf 34}, 2600 (1986).
\bibitem{wooters}
W. K. Wooters and W. H. Zurek, {\em Phys. Rev. D} {\bf 19}, 473 (1979).
\bibitem{greenberger}
D. M. Greenberger and A. Yasin, {\em Phys. Lett. A} {\bf 128}, 391 (1988).
\bibitem{englert}
B. -G. Englert, {\em Phys. Rev. Lett.} {\bf 77}, 2154-2157 (1996).
\bibitem{ghose b}
P. Ghose, {\em Testing quantum mechanics on new ground}, Cambridge University Press (1999), Chapter 1.
\bibitem{dbb}
D. Bohm, {\em Phys. Rev.} {\bf 85}, 166-93 (1952).
\bibitem{gha}
P. Ghose, D. Home and G. S. Agarwal, {\em Phys. Lett. A} {\bf 153}, 403-06 (1991); {\bf 168}, 95-99 (1992).
\bibitem{gh}
P. Ghose and D. Home, {\em Foundations of Physics} {\bf 22}, 1435-14 (1992).
\bibitem{mo}
Y. Mizobuchi and Y. Ohtak\'{e}, {\em Phys. Lett. A} {\bf 168}, 1 (1992).
\bibitem{brida}
G. Brida, M. Genovese, M. Gramegna and E. Predazzi, {\em Phys. Lett. A} {\bf 328}, 313 (2004).
\bibitem{vigier}
N. C. Petroni and Jean-Pierre Vigier, {\em Found. of Phys. Lett.} {\bf 14}, 395-400 (2001). 
\bibitem{nelson}
E. Nelson, {\em  Phys. Rev.} {\bf 150}, 1079-1085 (1996)..
\bibitem{schr}
E. Schr\"{o}dinger, {\em Naturwissenschaften} {\bf 23}, 807-812; 823-828; 844-949 (1935). The first English translation appeared in {\em Proc. Am. Phil. Soc} {\bf 124}, 323-38 (1980).
\bibitem{leg1}
A. O. Caldeira and A. J. Leggett, {\em Phys. Rev. Lett.} {\bf 46}, 211--14 (1981).
\bibitem{leg2}
A. J. Leggett, S. Chakravarty, A. Y. Dorsey, M. P. A. Fisher, A. Garg and W. Zwerger, {\em Rev. Mod. Phys.} {\bf59}, 1-85 (1987).
\bibitem{leg3}
U. Weiss, H. Grabert and S. Linkwitz, {\em J. Low Temp. Phys.} {\bf 68}, 213-244 (1987).
\bibitem{squid}
J. R. Friedman, V. Patel, W. Chen, S. K. Tolpygo and J. E. Lukens, {\em nature} {\bf 406}, 43-46 (2000).
\bibitem{epr}
A. Einstein, B. Podolski and N. Rosen, {\em Phys. Rev.} {\bf 47}, 777-80 (1035).
\bibitem{ein}
A. Einstein, ``Autobiographical Notes'', {\em Albert Einstein: Philosopher Scientist}, P. A. Schilpp (ed.), Harper \& Row Pub., New York and London (1059 edition), p. 85.
\bibitem{bac} 
G. Bacciagaluppi, A. Valentini: {\em Quantum Theory at the Crossroads: Reconsidering the 1927 Solvay Conference}, pp. 485-487, Cambridge Univ. Press, Cambridge (2009); arXiv: quant-ph/0609184.
\bibitem{bell1} 
J. S. Bell, {\em Physics} {\bf 1}, 1 (1964).
\bibitem{vN2}
J. von Neumann, {\em Mathematical Foundations of Quantum Mechanics}, Princeton University Press, Princeton (1955); original German version was published in 1932. 
\bibitem{ev}
H. Everett III, {\em Rev. Mod. Phys.} {\bf 29}, 454-62 (1957).
\bibitem{dewitt}
B. S. M. De Witt, {\em Phys. Today} {\bf 23}, 30-35 (1970).
\bibitem{hargreaves}
H. Hargreaves, `Probability in the Everett Interpretation', http://users.ox.ac.uk/~mert2255/papers/pitei.pdf, and references therein.
\bibitem{adlam}
E. Adlam, {\em Studies in Hist. and Phil. of Sc. Part B: Studies in Hist. and Phil. of Mod. Phys.} {\bf 47}, 21-32 (2014). 
\bibitem{dawid}
R. Dawid and K. P. Y. Th\'{e}bault, {\em Studies in Hist. and Phil. of Sc. Part B: Studies in Hist. and Phil. of Mod. Phys.} {\bf 47}, 55-61 (2014).
\bibitem{teg}
A. Aguirre and M. Tegmark, {\em Phys. Rev. D} {\bf 84}, 105002; arXiv:1008.1066.
\bibitem{gen2}
M. Genovese, {\em Phys. Reports} {\bf 413}, 319-96 (2005).
\bibitem{hiley}
D. Bohm and and B. J. Hiley, {\em The Undivided Universe: An Ontological Interpretation of Quantum Theory}, Routlege (1993).
\bibitem{eins2}
{\em Born-Einstein Letters}, Macmillan Press, 1971, p.192.
\bibitem{holland}
P. R. Holland, {The Quantum Theory of Motion}, Cambridge University Press (1993).
\bibitem{durr}
D. D\"{u}rr and S. Teufel, {\em Bohmian Mechanics: The Physics and Mathematics of Quantum Theory}, Berlin: Springer-Verlag (2009).\\
D. D\"{u}rr, S. Goldstein and N. Zanghi, {\em J. of Stat. Phys.} {\bf 67}: 843-907 (1992); {\bf 68}, 259-270 (1992).
\bibitem{valentini}
A. Valentini, {\em Phys. Lett. A} {\bf 158}, 1-8 (1991).
\bibitem{ghose3}
P. Ghose, {\em Adv. Sc. Letters} {\bf 2}, 97-99 (2009).
\bibitem{marco2}
M. Genovese, G. Brida, M. Gramegna, F. Piacentini, E. Predazzi and I. Ruo-Berchera, {\em J. Phys, : Conference Series}
{\bf 67}, 012047 (2007); arxiv: 061207 {quant-ph}, (2006).
\bibitem{kemmer}
N. Kemmer, {\em Proc. Roy. Soc.} {\bf 173}, 91-116 (1939).
\bibitem{hc}
Harish-Chandra, {\em Proc. Roy. Soc.} {\bf 186}, 502-525 (1946). 
\bibitem{ghosehome}
P. Ghose, D. Home and M. N. Sinha Roy, {\em Phys. Lett.  A} {\bf 183}, 267-271 (1993) ; {\em A 188}, 402 (Erratum).\\
P. Ghose, D. Home,  {\em Phys. Lett.  A} {\bf 191}, 362-364 (1994).\\
P. Ghose, {\em Found. of Phys.} {\bf 26}, 1441-1455 (1996).
\bibitem{gmg}
P. Ghose, A. S. Majumdar, S. Guha and J. Sau, {\em Phys. Lett. A} {\bf 290}, 205-213 (2001).
\bibitem{steinberg}
X. Xing, S. Kocsis, S. Ravets, L. K. Shalm, Y. Soudagar, F. Wolfgramm,
M. W. Mitchell, M. J. Stevens, R. P. Mirin and A. M. Steinberg, {\em Towards fundamental tests and quantum information aplications using novel photon sources}, Single Photon Workshop 2009, Boulder.\\
S. Kocsis, S. Ravets, B. Braverman, L. Krister Shalm, A. M. Steinberg: {\em Observing the Trajectories of a Single Photon Using Weak Measurement}, 19th Australian Institute of Physics (AIP) Congress, (2010).\\
S. Kocsis, B. Braverman, S. Ravets, M. J. Stevens, R. P. Mirin, L. Krister Shalm, A. M. Steinberg, {\em Science} {\bf 332}, 1170-1173 (2011).
\bibitem{weak}
Y. Aharonov, D. Z. Albert and L, Vaidman, {\em Phys. Rev. Lett.} {\bf 60}, 1351-1354 (1988).\\
Y. Aharonov and L. Vaidman, `The Two-State Vector Formalism of Quantum Mechanics: an Updated Review', {\em Lecture Notes in Physics} {\bf 734}, 399-447 (2007). arXiv:quant-ph/0105101.\\
B. E. Y. Svensson, {\em Quanta} {\bf 2} (1), 18-49 (2013).
\bibitem{cont}
A. Peres, {\em Phys. Rev. Lett.} {\bf 62}, 2326 (1989).\\ 
I. M. Duck and P. M. Stevenson, {\em Phys. Rev. D} {\bf 40}, 2112-2117 (1989).\\
S. Parrott, arXiv:0909.0295v3 [quant-ph] Jan 2010.
\bibitem{griffiths}
R. B. Griffiths, {\em Consistent Quantum Theory}, Cambridge University Press, 2003.\\
R. Omnès, {\em Quantum Philosophy}, Princeton University Press, 1999.
\bibitem{cramer}
J. Cramer, {\em Rev. Mod. Phys.} {\bf 58}, 647-688, July (1986).
\bibitem{modal}
B. C. van Fraassen, ``A formal approach to the philosophy of science'' in {\em Paradigms and Paradoxes: The Philosophical Challenge of the Quantum Domain}, R. Colodny (ed.), Pittsburgh: University of Pittsburgh Press, pp. 303-366, (1972).\\
D. Dieks and P. Vermaas (eds.), {\em The Modal Interpretation of Quantum Mechanics}, Dordrecht: Kluwer Academic Publishers (1998).
\bibitem{rov}
C. Rovelli, {\em Int. J. Theoret. Phys.} {\bf35}, 1637-1678 (1996); arXiv:quant-ph/9609002
\bibitem{fuchs}
C. A. Fuchs, N. D. Mermin and R. Schack, {\em An Introduction to QBism with an Application to the Locality of Quantum Mechanics}, arXiv: 1131.5253 [quant-ph] Nov 2013.
\bibitem{bell2}
J. S. Bell, {\em Rev. Mod. Phys.} {\bf 38}, 447-52 (1966).
\bibitem{bert}
R. A. Bertlmann, {\em J. Phys. A: Math. Theor.} {\bf 47}, 424007 (2014).
\bibitem{brown}
H. Brown and C. G. Timpson, {\em Bell on Bell's theorem: The changing face of nonlocality}, arxiv:1501.03521 [quant-ph] Dec 2014. 
\bibitem{zuk1}
M. Zukowski and C. Brukner, {\em Quantum non-locality - It ain’t necessarily so...}, arxiv: 1501.04618 [quant-ph] Jan 2015. 
\bibitem{zuk2}
M. Zukowski, {\em 
’s theorem tells us not what quantum mechanics is, but what quantum mechanics is not}, arxiv: 1501.05640 [quant-ph] Jan 2015. 
\bibitem{ks}
S. Kochen and E. Specker, {\em J. Math. and Mech.} {\bf 17}, 59-87 (1967).
\bibitem{mermin}
N. D. Mermin,  {\em Phys. Rev. Lett.} {\bf 65}, 3373 (1990); {\em Rev. Mod. Phys.} {\bf 65}, 803 (1993).
\bibitem{peres}
A. Peres, {\em Phys. Lett. A} {\bf 151}, 107 (1990); {\em J. Phys. A: Math. Gen.} {\bf 24}, L175-L178 (1991).
\bibitem{hasegawa}
Y. Hasegawa, K. Durstberger-Rennhofer, S. Sponar, and H. Rauch, {\em Nucl Instrum Methods Phys Res A.} {\bf 634}(1), S21-S24 (2011).
\bibitem{leg}
A. J. Leggett and A. Garg, {\em Phys. Rev. Lett.} {\bf 54}, 857 (1985). 
\bibitem{emary}
C. Emary, N. Lambert and F. Neri, {\em Leggett-Garg Inequalities}, arxiv: 1304.5133v3 [quant-ph] Jan 2014.
\bibitem{qi}
E. T. Jaynes, ``Probability in quantum theory'', in {\em Complexity, Entropy and the Physics of Information}. W.H. Zurek (ed), Addison Wesley Publishing (1990).
\bibitem{spekkens}
R.W. Spekkens, {\em Phys. Rev. A} {\bf 75}, 032110 (2007).
\bibitem{pbr}
M.\ F.\ Pusey, J.\ Barrett, and T.\ Rudolph, {\em Nature Phys.} {\bf 8}, 476 (2012).
\bibitem{cr}
R.\ Collbeck and R.\ Renner, {\em Phys. Rev. Lett.} {\bf 108}, 150402 (2012).
\bibitem{pg}
 P.\ G.\ Lewis, D.\ Jennings, J.\ Barrett, and T.\ Rudolph, {\em Phys. Rev. Lett.} {\bf 109}, 150404 (2012).
\bibitem{har}
N. Harrigan and W. R. Spekkens, {\em Found. Phys.} {\bf 40}, 125-157 (2010).
\bibitem{ghirardi}
G. C. Ghirardi and R. Romano, {\em Found. of Phys.}, {\bf 43}, 881 (2013); {\em Phys. Rev. A} {\bf 85}, 042102 (2012); {\em Phys. Rev. Lett.} {\bf 110}, 170404 (2013); arxiv: 1501.04127 [quant-ph] Jan 2015.
\bibitem{leifer}
M. S. Leifer, {\em Is the quantum state real? A review of $\psi$-ontology theorems}, arXiv: 1409.1570 [quant-ph] Sep 2014 and references therein.
\bibitem{kvn1}
E. Gozzi, {\em Phys. Lett. B} {\bf 201}, 525 (1988).
\bibitem{kvn2}
E. Gozzi, M. Reuter, and W. Thacker, {\em Phys. Rev. D} {\bf 40},
3363 (1989).
\bibitem{kvn3}
J. Wilkie and P. Brumer, {\em Phys. Rev. A} {\bf 55}, 27 (1997).
\bibitem{kvn4}
J. Wilkie and P. Brumer, {\em Phys. Rev. A} {\bf 55}, 43 (1997).
\bibitem{kvn5}
E. Gozzi and D. Mauro, {\em Ann. Phys.} {\bf 296}, 152 (2002).
\bibitem{mauro}
D. Mauro, Ph. D. thesis, arXiv:quant-ph/0301172.
\bibitem{kvn6}
E. Deotto, E. Gozzi, and D. Mauro, {\em J. Math. Phys.} {\bf 44},
5902 (2003).
\bibitem{kvn7}
E. Deotto, E. Gozzi, and D. Mauro, {\em J. Math. Phys.} {\bf 44},
5937 (2003).
\bibitem{kvn8}
A. A. Abrikosov, E. Gozzi, and D. Mauro, {\em Ann. Phys.}
{\bf 317}, 24 (2005).
\bibitem{kvn9}
M. Blasone, P. Jizba, and H. Kleinert, {\em Phys. Rev. A} {\bf 71},
052507 (2005).
\bibitem{kvn10}
P. Brumer and J. Gong, {\em Phys. Rev. A} {\bf 73}, 052109 (2006).
\bibitem{kvn11}
P. Carta, E. Gozzi, and D. Mauro, {\em Ann. Phys. (Leipzig)}
{\bf 15}, 177 (2006).
\bibitem{kvn12}
E. Gozzi and C. Pagani, {\em Phys. Rev. Lett.} {\bf 105}, 150604 (2010).
\bibitem{kvn13}
E. Gozzi and R. Penco, {\em Ann. Phys.} {\bf 326}, 876 (2011).
\bibitem{kvn14}
E. Cattaruzza, E. Gozzi, and A. F. Neto, {\em Ann. Phys.}
{\bf 326}, 2377 (2011).
\bibitem{schmidt}
E. Schmidt, {\em Math. Ann.} {\bf 63}, 433 (1907).
\bibitem{sud2}
T. N. Sherry and E. C. G. Sudarshan, {\em Phys. Rev. D} {\bf 18}, 4580 (1978).
\bibitem{sud3}
T. N. Sherry and E. C. G. Sudarshan, {\em Phys. Rev. D} {\bf 20}, 857 (1979).
\bibitem{sud4}
S. R. Gautam, T. N. Sherry and E. C. G. Sudarshan, {\em Phys. Rev. D} {\bf 20}, 3081 (1979).
\bibitem{sud5}
E. C. G. Sudarshan, arxiv: 0402134 (2004).
\bibitem{zurek}
W. H. Zurek, {\em Phys. Rev. D} {\bf 24}, 1516-1525 (1981); {\bf 26} 1862-1880 (1982).\\ {\em Rev. Mod. Phys.} {\bf 75}, 715-775 (2003).
\bibitem{zurek2}
W. H. Zurek and J.-P. Paz, {\em Phys. Rev. Lett.} {\bf 72}, 2508-2511 (1994).
\bibitem{zurek3}
W. H. Zurek, {\em Decoherence and the Transition from Quantum to Classical--Revisited}, Los Alamos Science {\bf 27}, 2-25 (2002).
\bibitem{zurek4}
W. H. Zurek, {\em "Quantum Darwinism, classical reality, and the randomness of quantum jumps"}, {\em Phys. Today} {\bf 67}, 44-50 (2014).
\bibitem{zeh}
H. D. Zeh in {\em Decoherence and the Appearance of a Classical World in Quantum Theory}, eds. D. Giulini, E. Joos, C. Kiefer, J. Kupsch, I. -O. Stamatescu and H. D. Zeh, Berlin: Springer (1996); second revised edition (2003).
\bibitem{adler}
S. L. Adler, `Why Decoherence has not Solved the Measurement Problem: A Response to P. W. Anderson', {\em Studies in Hist. and Phil. of Mod. Phys.} {\bf 34B}, 135-142 (2003).
\bibitem{bac2}
G. Bacciagaluppi, {\em The Role of Decoherence in Quantum Mechanics}, Stanford Encyclopedia of Philosophy, http://plato.stanford.edu/entries/qm-decoherence/ (2012).
\end{thebibliography}
\end{document}